\documentclass[a4paper,11pt]{article}
\pdfoutput=1 % if your are submitting a pdflatex (i.e. if you have
\usepackage{jheppub} % for details on the use of the package, please
\usepackage[T1]{fontenc} % if needed

\newcommand{\Rmnum}[1]{\expandafter\@slowromancap\romannumeral #1@}
\makeatother
\definecolor{lightgray}{rgb}{.7,.7,.7}

\definecolor{red}{rgb}{1,0,0}

\definecolor{blue}{rgb}{0,0,1}

\title{\boldmath Dynamical Stability of the Holographic System with Two Competing Orders}

\author[a]{Yiqiang Du,}
\author[b]{Shan-Quan Lan,}
\author[a,c]{Yu Tian,}
\author[b,d]{Hongbao Zhang}

\affiliation[a]{School of Physics, University of Chinese Academy of Sciences, Beijing 100049, China}
\affiliation[b]{Department of Physics, Beijing Normal University, Beijing 100875, China}
\affiliation[c]{State Key Laboratory of Theoretical Physics, Institute of Theoretical Physics, Chinese Academy of Sciences, Beijing 100190, China}
\affiliation[d]{Theoretische Natuurkunde, Vrije Universiteit Brussel, and The International Solvay Institutes,\\Pleinlaan 2, B-1050 Brussels, Belgium}

\emailAdd{duyiqiang12@mails.ucas.ac.cn}
\emailAdd{shanquanlan@mail.bnu.edu.cn}
\emailAdd{ytian@ucas.ac.cn}
\emailAdd{hzhang@vub.ac.be}

\abstract{We investigate the dynamical stability of the holographic system with two order parameters, which exhibits competition and coexistence of condensations. In the linear regime, we have developed the gauge dependent formalism to calculate the quasi-normal modes by gauge fixing, which turns out be considerably convenient. Furthermore, by giving different  Gaussian wave packets as perturbations at the initial time, we numerically evolve the full nonlinear system until it arrives at the final equilibrium state. Our results show that the dynamical stability is consistent with the thermodynamical stability. Interestingly, the dynamical evolution, as well as the quasi-normal modes, shows that the relaxation time of this model is generically much longer than the simplest holographic system. We also find that the late time behavior can be well captured by the lowest lying quasi-normal modes except for the non-vanishing order towards the single ordered phase.
To our knowledge, this exception is the first counter example to the general belief that the late time behavior towards a final stable state can be captured by the lowest lying quasi-normal modes. In particular, a double relation is found for this exception in certain cases.
}

\begin{document}
\maketitle
\flushbottom
\section{Introduction}
The correlated stability, namely, the equivalence between thermodynamical and dynamical stability in gravitational physics, is a long-standing problem and has recently received renewed attention due to AdS/CFT correspondence. AdS/CFT correspondence states that a gravitational system in the AdS bulk is dual to an ordinary laboratory type system without gravity on the boundary. In particular, the bulk black hole is dual to the boundary system at finite temperature. So it is tempting to conjecture that the correlated stability should hold for such holographic gravitational systems\cite{gub,gub1}. Actually there are plenty of holographic models available for such a test, where a lot of works have been done\cite{he1,he2,he3,he4,he5,he6,h3}.

Of special interests, the authors of \cite{he} investigated a holographic model with two charged scalar fields {not coupled directly to each other}, which are correlated to two order parameters in the dual field theory, in the four dimensional bulk AdS spacetime under the probe limit, which we call the BHMRS model. More than one order parameters may describe a more complicated system, which has been extensively studied in condensed matter physics (e.g., in multicomponent superconductors and superfluids) \cite{le1,le3,julien}.

In \cite{he}, a new superconducting phase with two coexisting orders was found in a certain range of chemical potential in the dual theory, besides the ordinary phase and the phase with only one superconducting order. The coexisting phase is competitive between the two order parameters, or in other words, when one of the two scalar fields' condensation increases, the condensation of the other would decrease. Taking account of the backreaction, the BHMRS model was studied in detail in \cite{lili} and a much richer phase structure was found, which is determined by calculating and minimizing the free energy (grand potential) of each phase when there are more than one possible phases for a fixed set of thermodynamic parameters.

In this paper, we focus on the dynamical (in)stability of the BHMRS model in the probe limit. In order for comparison with the thermodynamical (in)stability, we first calculate the free energy of the system in the rigorous probe limit. Then, we investigate the quasi-normal modes (QNMs) of the system. Since the QNM investigation only involves linearized equations of motion, it is a quick and reliable way to find out the phase structure of a model.

Actually, we propose a new method to calculate QNMs without using the gauge invariant forms in the coexisting phase. Our new method looks more convenient than other widely used methods, e.g., the determinant method proposed by \cite{amado} for which one should first find a maximal set of linearly independent numerical solutions of the linearized equations of motion. In particular, compared with the gauge invariant formalism, the boundary conditions and gauge fixing of the linearized equations of motion are treated in a very natural way in our method, which is remarkably adapted for comparison with the late time behaviors of the real time evolution.

In addition to describing the late time behaviors of the evolution, QNM can also be used to investigate the phase transitions. Although we do not show the figures in this paper, we have made sure that there will be the lowest lying quasi-normal mode passing through the real axis around the critical values where a stable phase converts to be unstable.

We investigate the real time dynamical (in)stability of the system by giving the system various strength of Gaussian wave packets as perturbations to different equilibrium states in the initial time and then numerically evolving the system to find out which phase is dynamically stable when there are more than one possible phases. We find that the phase whose free energy is the lowest in certain area of charge density is also dynamically stable, i.e., the dynamical stability is consistent with the thermodynamic stability in this holographic system.

Interestingly, we find that the relaxation time of the system is generically much longer than the simplest holographic system, both in the case of far from criticality. Actually, we first observe this phenomenon in the dynamical evolution, and later confirm it through careful identification of the corresponding QNM, which would otherwise be confused with the mode $w=0$. It needs further investigation to see if there is some physical mechanism behind this phenomenon.

We also investigate the late time behaviors of the above evolution, in comparison with the QNMs as linear perturbations over the final stable configuration, and find the late time behaviors of the coexisting phase and the order parameter that will decay to zero in non-coexisting phases can be described well by the QNMs. Unexpectedly, the late time behaviors of the order parameters that will condense in non-coexisting phases are not consistent with their own QNMs, which is believed to be due to the nonlinear effects, which we will discuss in some detail.

This paper is organized as follows. In the next section, we will review the aforementioned holographic model for our two band superconductor. Then in Section $3$ we confirm the phase diagram presented in \cite{he} by comparing the free energy for all possible static solutions in the canonical ensemble. As a warm-up,  in the subsequent section we develop the gauge dependent formalism with the gauge fixed to work out the quasi-normal modes, which indicates the thermodynamically stable phases are also dynamically stable at the linear level. In Section $5$, by evolving our system with different Gaussian wave packets as perturbations at the initial time we find the thermodynamically stable phases are  dynamically stable at the fully non-linear level. Moreover, it is found that the late time behavior can be well captured by the lowest lying quasi-normal modes except for the non-vanishing order parameters towards the single ordered phases. The reason for this exception comes essentially from the non-linear effect, or put it in a physical way, the competition between the two bands such that the relaxation time for the non-vanishing order is generically elongated by the dying-out order. We conclude our paper in the end.

\section{The Holographic Model}
We will consider a holographic superconductor model in (3+1) dimensional AdS. The action is given by\cite{he}
\begin{equation}
 S=\frac{1}{2\kappa^2}\int_M d^4x\sqrt{-g}(R+\frac{6}{L^2})+\int_Mdx^4\sqrt{-g}\mathcal{L}_{matt}.
\end{equation}
Here $L$ is the radius of AdS,  $\kappa^2=8\pi G$ with $G$ the gravitational constant in the bulk, and
the matter field Lagrangian  is given by
\begin{equation}\label{ho1}
\mathcal{L}_{matt}=\frac{1}{e_2^2}(-\frac{1}{4}F^{ab}F_{ab}-|D_1\psi_1|^2-|D_2\psi_2|^2-m_1^2|\psi_1|^2-m_2^2|\psi_2|^2)
\end{equation}
where $D_1=\nabla-i\frac{e_1}{e_2}A, D_2=\nabla-iA$ with $e_i$ and $m_i$ ($i=1,2$) the charge and mass of the scalar field $\psi_i$ ($i=1,2$), respectively.

In this paper we shall work with the probe limit in which the back reaction of matter fields to the metric can be ignored by taking $e_2\rightarrow\infty$ but keeping the ratio $e=e_1/e_2$ finite. This allows us to start with the planar black hole as follows
\begin{equation}
 ds^2=\frac{L^2}{z^2}(-f(z)dt^2-2dtdz+dx^2+dy^2),
\end{equation}
where $ f(z)=1-(\frac{z}{z_h})^3$ with $z=0$ the AdS conformal boundary and $z=z_h$ the position of horizon. The temperature of the dual boundary system is given by the Hawking temperature
\begin{equation}
T=\frac{3}{4\pi z_h}.
\end{equation}
Then on top of this background geometry the probe matter fields are governed by the following equations of motion, i.e.,
\begin{align*}
D_{1a}D_1^a\psi_1-m_1^2\psi_1=0,\quad D_{2a}D_2^a\psi_2-m_2^2\psi_2=0,
\end{align*}
\begin{eqnarray}\label{eom}
 \nabla_aF^{ab}=j^b=ie[\psi_1^*D_1^b\psi_1-\psi_1(D_1^b\psi_1)^*]+i[\psi_2^*D_2^b\psi_2-\psi_2(D_2^b\psi_2)^*].
\end{eqnarray}

For simplicity but without loss of generality, below we will focus exclusively on the case of $m_1^2L^2=0, m_2^2L^2=-2$. Moreover, to make our life easier, we shall turn off $A_x, A_y$, and require all the other quantities depend only on coordinates $t$ and $z$.
With this, the asymptotic behavior of the solutions near the AdS boundary can be written as
\begin{equation*}
\psi_1=\frac{1}{L}(\psi_{-,1}+\psi_{+,1}z^3+\dots),
\end{equation*}
\begin{equation*}
\psi_2=\frac{1}{L}(\psi_{-,2}z+\psi_{+,2}z^2+\dots),
\end{equation*}
\begin{eqnarray}
A_t=\mu-\rho z+\dots,
\end{eqnarray}
where we have been working in the axial gauge $A_z=0$.
According to the usual dictionary in the AdS/CFT correspondence, we know that the constants $\mu$ and $\rho$ can be regarded as the chemical potential and charge density of the dual boundary theory, respectively. In addition, when the sources $\psi_{-,1}$ and $\psi_{-,2}$ are switched off, $\psi_{+,1}$ and $\psi_{+,2}$ correspond to the spontaneous broken order parameters of our two band holographic superconductor with dimension three and dimension two, respectively.

In what follows, we will work in the units where $L=1$ and $z_h=1$, making use of the scaling symmetry of AdS. Then the evolution equations can be written as
\begin{eqnarray}\label{psi}
\frac{f}{2}\psi^{\prime\prime}_1-(\partial_t-ieA_t)\psi^{\prime}_1+\frac{1}{z}(\partial_t-ieA_t)\psi_1+\frac{ie}{2}A^{\prime}_t\psi_1+(\frac{f^{\prime}}{2}-\frac{f}{z})\psi^{\prime}_1=0,
\end{eqnarray}
\begin{eqnarray}\label{chi}
\frac{f}{2}\chi^{\prime\prime}-(\partial_t-iA_t)\chi^{\prime}+\frac{i}{2}A^{\prime}_t\chi+\frac{f^{\prime}}{2}\chi^{\prime}-\frac{z}{2}\chi=0,
\end{eqnarray}
\begin{align}\label{At}
\partial_tA_t^{\prime}=&\frac{ief}{z^2}(\psi_1^*\psi_1^{\prime}-\psi_1\psi_1^{*\prime})-\frac{1}{z^2}[ie(\psi_1^{*}\partial_t\psi_1-\psi_1\partial_t\psi_1^*)+2e^2\psi_1^*\psi_1A_t]\nonumber\\
&+if(\chi^*\chi^{\prime}-\chi\chi^{*\prime})-[i(\chi^{*}\partial_t\chi-\chi\partial_t\chi^*)+2\chi^*\chi A_t],
\end{align}
with the constraint equation
\begin{align}\label{constr}
A_t^{\prime\prime}=\frac{ie}{z^2}(\psi_1^*\psi_1^{\prime}-\psi_1\psi_1^{*\prime})+i(\chi^*\chi^{\prime}-\chi\chi^{*\prime}),
\end{align}
where $\chi=\frac{\psi_2}{z}$.

\section{Phase Diagram and Thermodynamical Stability}
In this section, we shall reproduce the phase diagram for the homogeneous and isotropic two band holographic superconductor. The corresponding bulk static equations of motion are
\begin{align}\label{ste1}
\frac{f}{2}\psi_1^{\prime\prime}+ieA_t\psi_1^{\prime}-\frac{ie}{z}A_t\psi_1+\frac{ie}{2}A_t^{\prime}\psi_1+(\frac{f^{\prime}}{2}-\frac{f}{z})\psi_1^{\prime}=0,
\end{align}
\begin{align}
\frac{f}{2}\chi^{\prime\prime}+iA_t\chi^{\prime}+\frac{i}{2}A_t^{\prime}\chi+\frac{f^{\prime}}{2}\chi^{\prime}-\frac{z}{2}\chi=0,
\end{align}
\begin{align}
\frac{ief}{z^2}(\psi_1^*\psi_1^{\prime}-\psi_1\psi_1^{*\prime})-\frac{2e^2}{z^2}\psi_1^*\psi_1A_t+if(\chi^*\chi^{\prime}-\chi\chi^{*\prime})-2\chi^*\chi A_t=0,
\end{align}
\begin{align}\label{ste2}
A_t^{\prime\prime}-\frac{ie}{z^2}(\psi_1^*\psi_1^{\prime}-\psi_1\psi_1^{*\prime})-i(\chi^*\chi^{\prime}-\chi\chi^{*\prime})=0.
\end{align}
There is a trivial solution $\psi_1=\chi=0$ and $A_t=\rho(1-z)$, which corresponds to the normal phase of our boundary system by holography. But generically it is hard to find out the analytic expression for the static bulk solution dual to the boundary superconducting phase. So we would like to obtain the corresponding numerical solution by pseudo-spectral method. To achieve this, we first
set $\psi_1=|\psi_1|e^{ie\theta_1}=\phi_1e^{ie\theta_1}$ and $\chi=|\chi|e^{i\theta_2}=\phi_2e^{i\theta_2}$, and then rewrite the above static equations to get the following five independent ones (see Appendix for details):
\begin{align}\label{static1}
f\phi_1^{\prime\prime}+f^{\prime}\phi_1^{\prime}-\frac{2f}{z}\phi_1^{\prime}-e^2(2A_t\theta_1^{\prime}+f\theta_1^{\prime2})\phi_1=0,
\end{align}
\begin{align}\label{static2}
f\phi_2^{\prime\prime}+f^{\prime}\phi_2^{\prime}-z\phi_2-(2A_t\theta_2^{\prime}+f\theta_2^{\prime2})\phi_2=0,
\end{align}
\begin{align}\label{static4}
A_t+f\theta_1^{\prime}=0.
\end{align}
\begin{align}\label{static5}
A_t+f\theta_2^{\prime}=0.
\end{align}
\begin{align}\label{static3}
A_t^{\prime\prime}+\frac{2e^2}{z^2}\theta_1^{\prime}\phi_1^2+2\theta_2^{\prime}\phi_2^2=0.
\end{align}

\begin{figure}[htbp]
\centering
\includegraphics[scale=0.38]{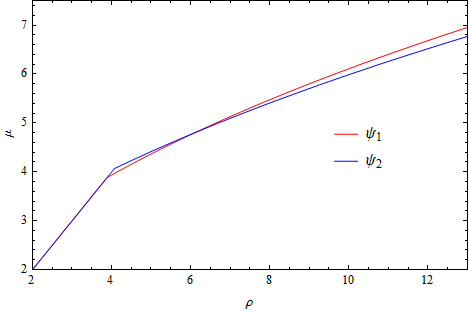}
\includegraphics[scale=0.48]{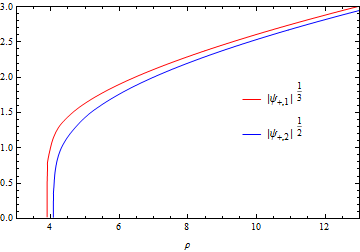}
\caption{The left panel shows the relation between chemical potential $\mu$ and charge density $\rho$ in the one band holographic superconductor while the right one shows the corresponding condensation with respect to the charge density, where the critical charge density is given by $\rho_1=3.891166$ and $\rho_2=4.0745$, respectively.}\label{fig1}
\end{figure}

\begin{figure}[htbp]
\centering
\includegraphics[scale=0.37]{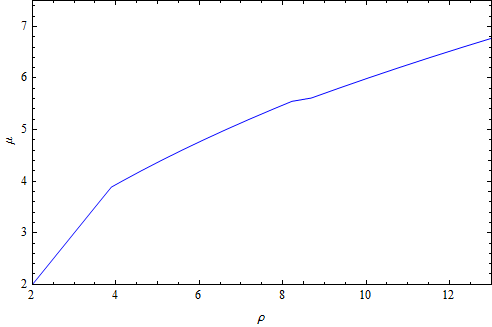}
\includegraphics[scale=0.49]{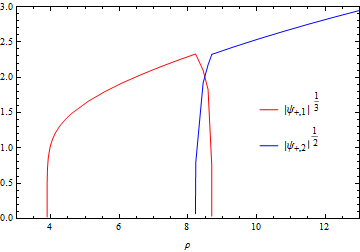}
\caption{The left panel shows the relation between chemical potential $\mu$ and charge density $\rho$ in our two band holographic superconductor while the right one shows the corresponding condensation with respect to the charge density, where the lower and upper critical charge density for the coexistent phase are given by $\rho_{c1}=8.21996$ and $\rho_{c2}=8.69349$, respectively.}\label{fig2}
\end{figure}

Therefore we shall solve the above equations of motion to obtain our numerical solution. To this end, we are required to specify the appropriate boundary conditions. $\theta_1=\theta_2=0$ can be achieved on the horizon by gauge fixing. In addition, the above equations of motion give rise to the following natural boundary conditions
\begin{align}
A_t=0,\qquad\phi_1^{\prime}=0,\qquad\phi_2+3\phi_2^{\prime}=0
\end{align}
on the horizon. Finally since we want to find out the spontaneously broken phase for our boundary system at a fixed charge density, we impose the boundary conditions on the AdS boundary as follows:
\begin{align}\label{e1}
\phi_1=0,\qquad\phi_2=0,\qquad A_t^{\prime}=-\rho.
\end{align}
When there is only $\psi_1$ under the condensation, we call the phase as Phase-I. Phase-II represents the phase in which only $\psi_2$ is under the condensation. By our numerics we obtain the phase diagram in Figure~\ref{fig1} for the one band holographic superconductor, and the phase diagram in Figure~\ref{fig2} for our two band holographic superconductor. Here the condensations of the two scalar fields can coexist in a certain range of $\rho$, which we call as Phase-III.

In Figure~\ref{fig1}, we denote the critical charge density for the condensations of scalar fields $\psi_1$ and $\psi_2$ as $\rho_1$ and $\rho_2$, respectively. In Figure~\ref{fig2}, the occurrence of the attenuation of the condensation of $\psi_1$ is also the beginning of the emergence of condensation of $\psi_2$, and the corresponding critical charge density is denoted as $\rho_{c1}$. We use $\rho_{c2}$ to denote the critical charge density where the condensation of $\psi_1$ vanishes.


To make sure Figure~\ref{fig2} represents the genuine phase diagram for our two band holographic superconductor, we are required to check whether the corresponding free energy density is the lowest among all possible phases\footnote{This has not been achieved in \cite{he}.}.
Upon taking into account the back reaction, \cite{lili} shows that the grand potential results in the case of weak back reaction are consistent with the claim in \cite{he}, but direct confirmation in the probe limit is still lacking. Now we will calculate the free energy in the real probe limit.
By holography, the free energy density can be obtained from the on shell Lagrangian of matter fields as follows:
\begin{align}
F=& -e_2^2 [\int dz\sqrt{-g}\mathcal{L}_{matt}+\sqrt{-h}(n_a A_b F^{ab} +|\partial\psi_1|^2-|\psi_2|^2)|_{z=0}]\nonumber\\
=&-\frac{1}{2}\int dz(\sqrt{-g}A_aj^a+\sqrt{-h}n_a A_b F^{ab}|_{z=0}).
\end{align}
In the first line we have added the first boundary term to guarantee that we are working in the canonical ensemble rather than the grand canonical ensemble, and the last two boundary terms as the counter terms in the holographic renormalization to make the bulk on shell action finite, where $h$ is the determinant of the induced metric on the AdS boundary. In the second line, we have made use of the equations of motion and the asymptotic AdS boundary behavior for the static solutions of matter fields.

\begin{figure}[htbp]
\centering
\includegraphics[scale=0.38]{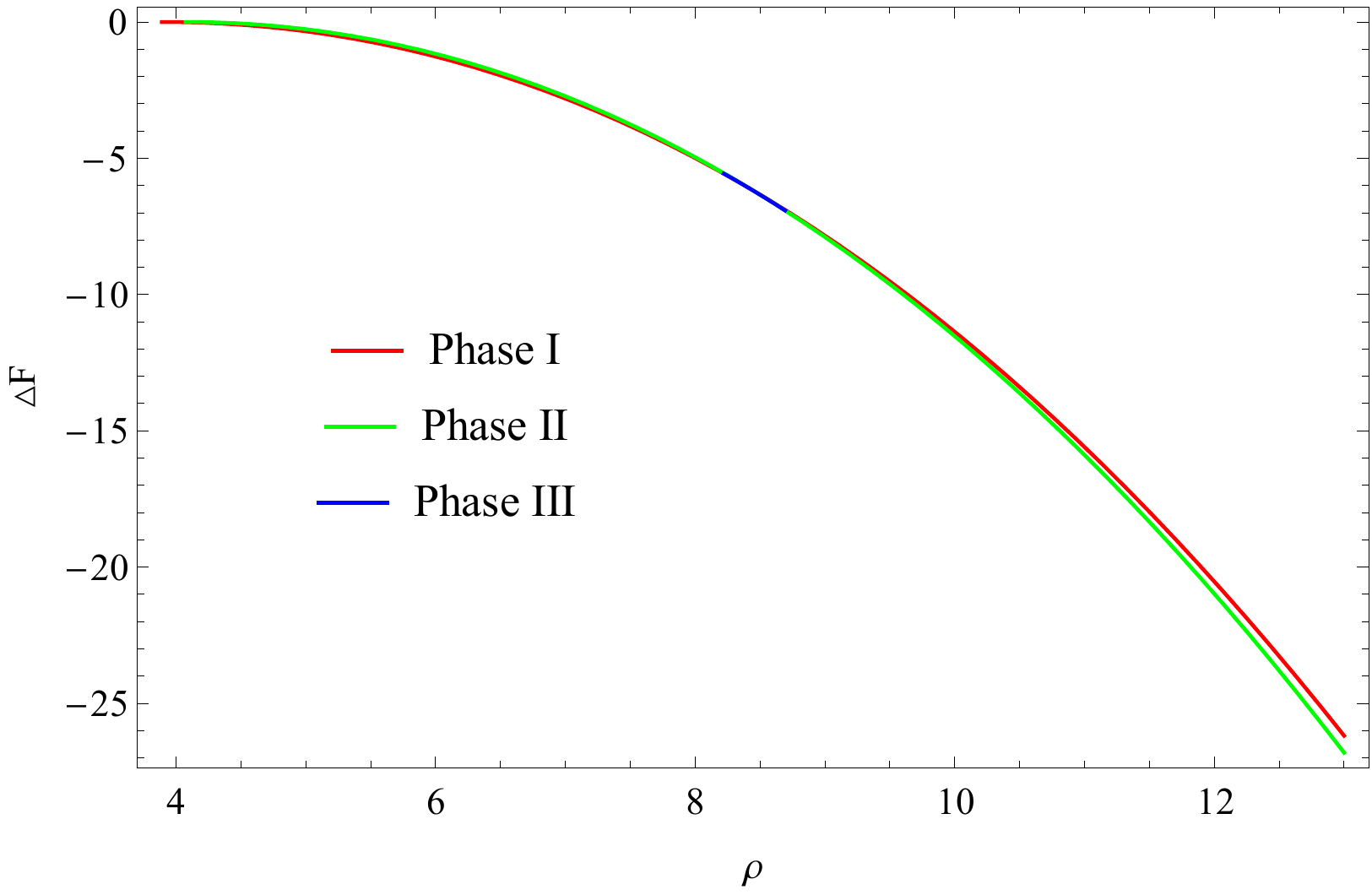}
\includegraphics[scale=0.41]{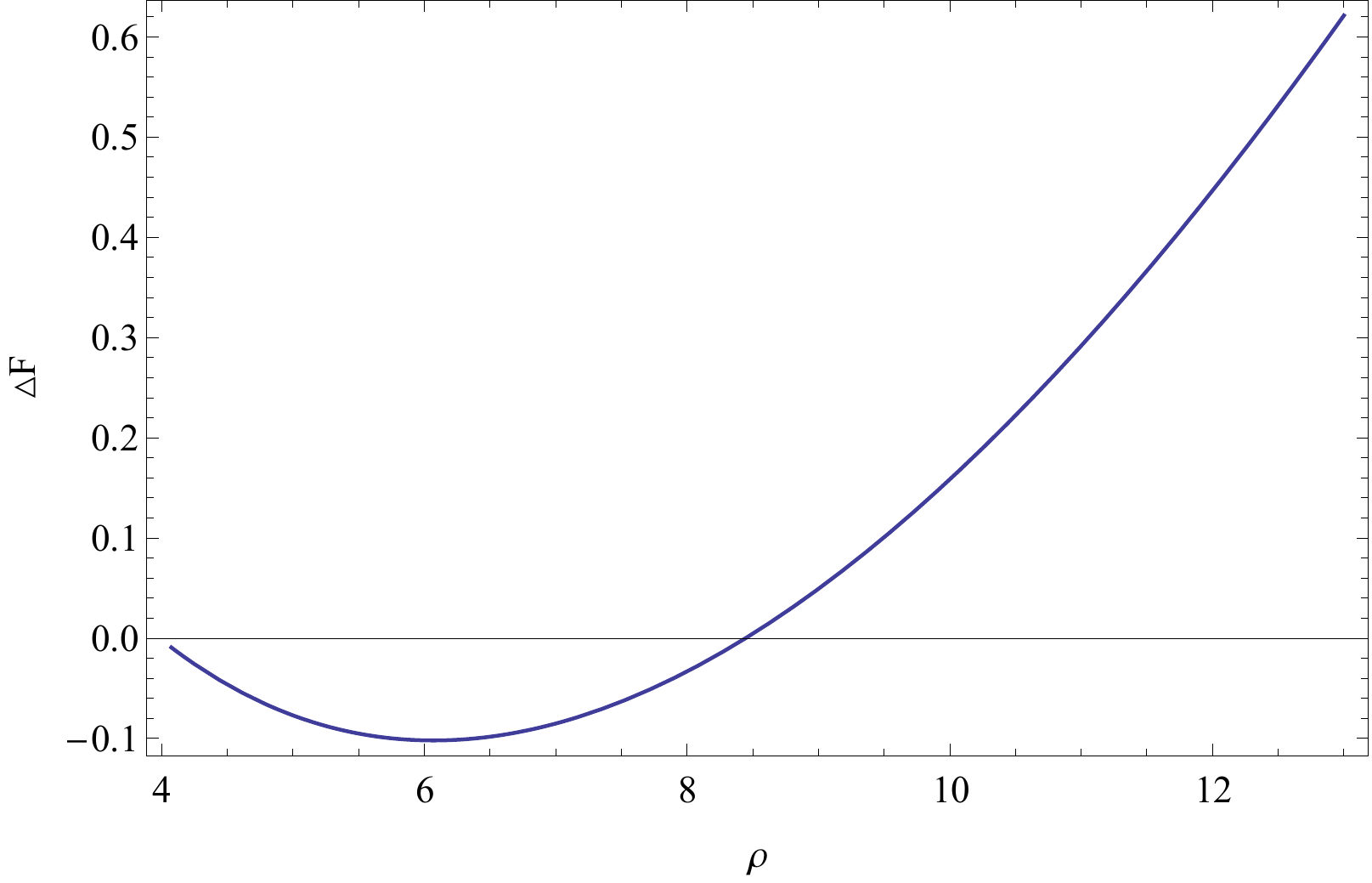}
\caption{The left panel shows the free energy density difference of the three superconducting phases from the normal phase. The right panel shows the free energy density difference of Phase-I from Phase-II. Whence Phase-I is thermodynamically stable when the charge density is between $\rho_1$ and  $\rho_{c1}$, while Phase-II is thermodynamically stable when the charge density is larger than $\rho_{c2}$.}\label{fre1}
\end{figure}

\begin{figure}[htbp]
\centering
\includegraphics[scale=0.375]{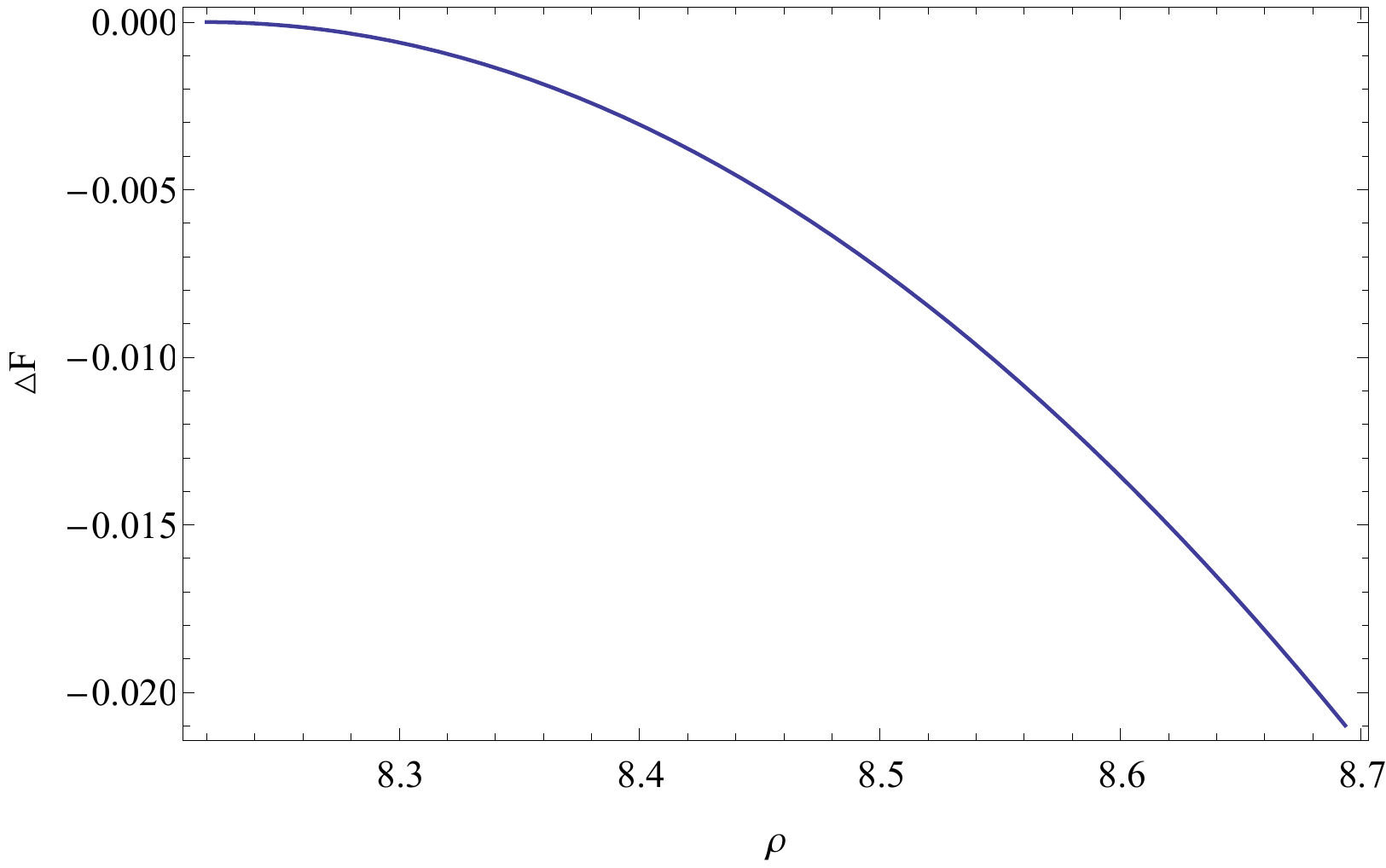}
\includegraphics[scale=0.39]{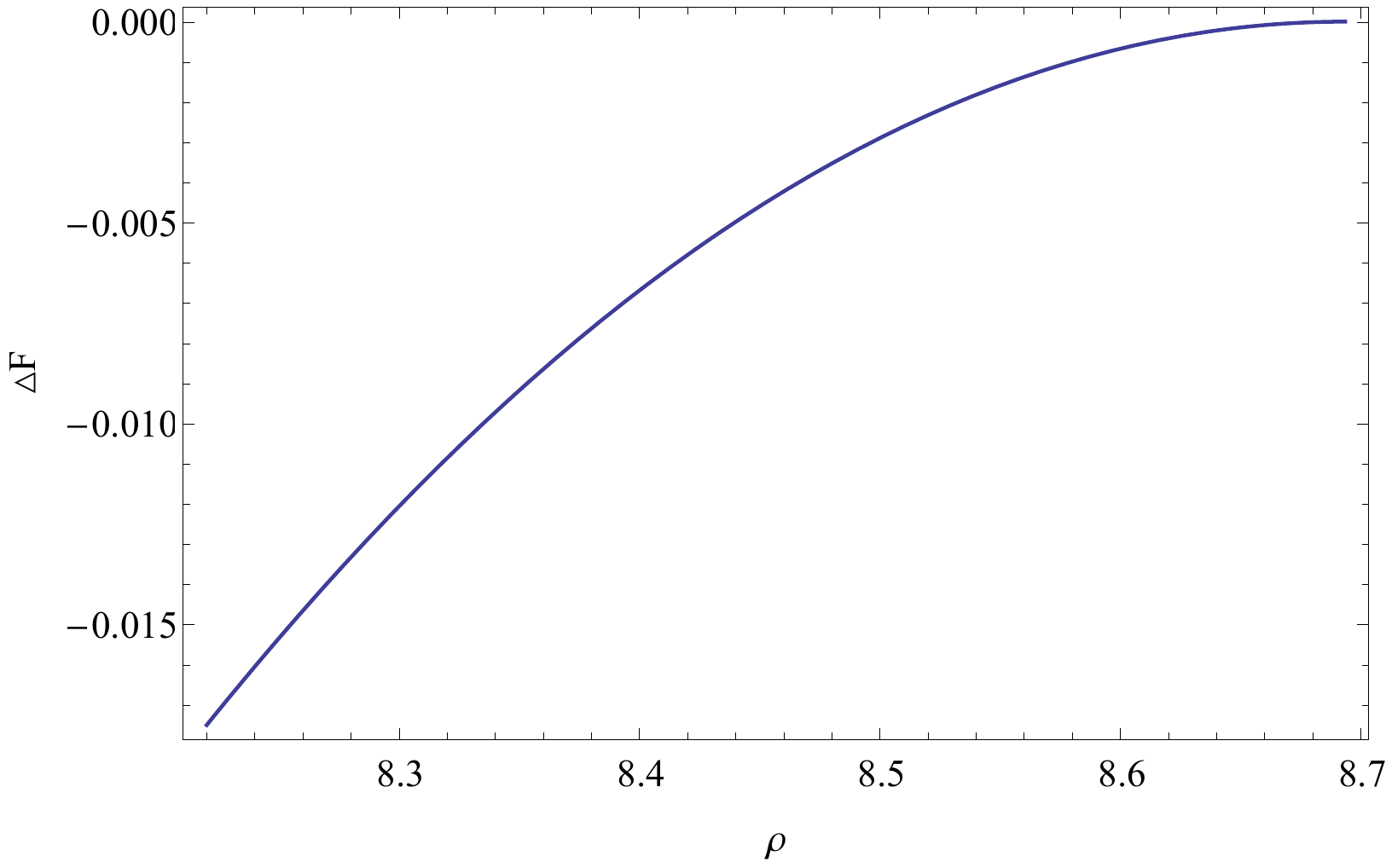}
\caption{The two panels show the free energy density difference of Phase-III from Phase-I and Phase-II between $\rho_{c1}$ and $\rho_{c2}$, respectively. Whence Phase-III is the thermodynamically stable phase in this region.}\label{fre2}
\end{figure}

As shown in the left panel of Figure~\ref{fre1}, compared to the normal phase the free energy density of three superconducting phases is much lower.  Furthermore, as demonstrated in Figure~\ref{fre2} and the right panel of Figure~\ref{fre1}, by scrutinizing the free energy density difference among these three superconducting phases, we lead to the above desired phase diagram in the canonical ensemble.


\section{Quasi-normal Modes}
Before we move onto the fully non-linear dynamical stability by the real time evolution of our bulk fields, we would like to pause to play a little bit with the quasi-normal modes for our holographic model as a warm-up because the spectrum of quasi-normal modes can be regarded as a diagnosis of dynamical stability in the linear regime. If the phase in consideration is dynamically stable, then all the quasi-normal modes sit in the lower complex frequency plane. Actually as we shall show later on,  all the thermodynamically stable phases in consideration is dynamically stable at such a linear level.  In addition, he onset of Goldstone mode at the origin generically signals a transition from one phase to another. Moreover, among others, the lowest lying QNM is believed to capture the late time behavior during the real time fully non-linear evolution towards the desired final state as the late time perturbation is supposed to be small enough to validate the linear perturbation theory.

To our knowledge, so far there have been two methods developed  to calculate the quasi-normal modes for the gauged systems. One is the gauge invariant formalism \cite{leaver}, and the other is the gauge dependent formalism, where the gauge degree of freedom is freed\cite{amado}. We would like to take a third route by the gauge dependent formalism but with the fixed gauge. As seen later on, this alternative method is highly efficient and particularly suited for the comparison with the late time behavior of fully non-linear dynamical evolution of bulk fields.

To begin with, we first write down the linear perturbation equations of \eqref{psi}, \eqref{chi} and \eqref{constr} as
\begin{align}\label{per1}
&-iw\delta\psi_1^{\prime}-ie_1\psi_1^{\prime}\delta A_t-ie_1A_t\delta\psi_1^{\prime}+\frac{iw}{z}\delta\psi_1+\frac{ie_1}{z}\psi_1\delta A_t\nonumber\\
&+\frac{ie_1}{z}A_t\delta\psi_1-\frac{ie_1}{2}\psi_1\delta A_t^{\prime}-\frac{ie_1}{2}A_t^{\prime}\delta\psi_1-\frac{f}{2}\delta\psi_1^{\prime\prime}-(\frac{f^{\prime}}{2}-\frac{f}{z})\delta\psi_1^{\prime}=0,
\end{align}
\begin{align}\label{per2}
&-iw\delta\chi^{\prime}-ie_2\chi^{\prime}\delta A_t-ie_2A_t\delta\chi^{\prime}-\frac{ie_2}{2}\chi\delta A_t^{\prime}\nonumber\\
&-\frac{ie_2}{2}A_t^{\prime}\delta\chi-\frac{f}{2}\delta\chi^{\prime\prime}-\frac{f^{\prime}}{2}\delta\chi^{\prime}+\frac{z}{2}\delta\chi=0,
\end{align}
\begin{align}\label{per3}
\delta A_t^{\prime\prime}=&\frac{ie_1}{z^2}(\delta\psi_1^*\psi_1^{\prime}+\psi_1^*\delta\psi_1^{\prime}-\delta\psi_1\psi_1^{*\prime}\nonumber\\
&-\psi_1\delta\psi_1^{*\prime})+ie_2(\delta\chi^*\chi^{\prime}+\chi^*\delta\chi^{\prime}-\delta\chi\chi^{*\prime}-\chi\delta\chi^{*\prime}).
\end{align}
Here we simply ignore the linear perturbation equation of \eqref{At} because it will be satisfied automatically in the whole bulk  once it holds at the AdS boundary by our later boundary condition there. To calculate the quasi-normal modes, we further make the following consistent ansatz for the perturbed fields, i.e.,
\begin{align}\label{delta}
&\delta\psi_1=p e^{-iwt}+\bar{p}e^{iw^*t},\nonumber\\
&\delta\chi=q e^{-iwt}+\bar{q}e^{iw^*t},\nonumber\\
&\delta A_t=\frac{1}{i}(a e^{-iwt}-a^{*}e^{iw^*t})
\end{align}
with $p$, $\bar{p}$, $q$, $\bar{q}$ and $a$ complex functions of $z$. It is noteworthy that $\bar{p}$ is independent of $p$ and $\bar{q}$ is independent of $q$.

Then substituting \eqref{delta} into the above perturbation equations \eqref{per1}-\eqref{per3}, we eventually end up with the following equations
\begin{align}\label{nper1}
&-iwp^{\prime}-e_1a\psi_1^{\prime}-ie_1A_tp^{\prime}+\frac{iw}{z}p+\frac{e_1}{z}a\psi_1+\frac{ie_1}{z}A_tp\nonumber\\
&-\frac{e_1}{2}a^{\prime}\psi_1-\frac{ie_1}{2}A_t^{\prime}p-\frac{f}{2}p^{\prime\prime}-(\frac{f^{\prime}}{2}-\frac{f}{z})p^{\prime}=0,
\end{align}
\begin{align}\label{nper2}
&-iw\bar{p}^{*\prime}+e_1a\psi_1^{*\prime}+ie_1A_t\bar{p}^{*\prime}+\frac{iw}{z}\bar{p}^*-\frac{e_1}{z}a\psi_1^*-\frac{ie_1}{z}A_t\bar{p}^*\nonumber\\
&+\frac{e_1}{2}a^{\prime}\psi_1^*+\frac{ie_1}{2}A_t^{\prime}\bar{p}^*-\frac{f}{2}\bar{p}^{*\prime\prime}-(\frac{f^{\prime}}{2}-\frac{f}{z})\bar{p}^{*\prime}=0,
\end{align}
\begin{equation}\label{nper3}
-iwq^{\prime}-e_2\chi^{\prime}a-ie_2A_tq^{\prime}-\frac{e_2}{2}a^{\prime}\chi-\frac{ie_2}{2}A_t^{\prime}q-\frac{f}{2}q^{\prime\prime}-\frac{f^{\prime}}{2}q^{\prime}+\frac{z}{2}q=0,
\end{equation}
\begin{equation}\label{nper4}
-iw\bar{q}^{*\prime}+e_2\chi^{*\prime}a+ie_2A_t\bar{q}^{*\prime}+\frac{e_2}{2}a^{\prime}\chi^*+\frac{ie_2}{2}A_t^{\prime}\bar{q}^*-\frac{f}{2}\bar{q}^{*\prime\prime}-\frac{f^{\prime}}{2}\bar{q}^{*\prime}+\frac{z}{2}\bar{q}^*=0,
\end{equation}
\begin{equation}\label{nper5}
a^{\prime\prime}+\frac{e_1}{z^2}(\bar{p}^*\psi_1^{\prime}+\psi_1^*p^{\prime}-p\psi_1^{*\prime}-\psi_1\bar{p}^{*\prime})+e_2(\bar{q}^*\chi^{\prime}+\chi^*q^{\prime}-q\chi^{*\prime}-\chi\bar{q}^{*\prime})=0
\end{equation}
 with
\begin{equation}\label{bdry}
p=\bar{p}^*=q=\bar{q}^*=a^{\prime}=0
\end{equation}
on the AdS boundary.

Note that for a small parameter $\lambda$ and arbitrary $w$ the gauge transformation
\begin{equation}
A\rightarrow A+\nabla\theta,\quad \psi_1\rightarrow e^{ie\theta}\psi_1, \quad \chi\rightarrow e^{i\theta}\chi
\end{equation}
with
\begin{equation}
\theta=\frac{1}{i}(\lambda e^{-iw t}-\lambda^* e^{iw^*t})
\end{equation}
always induces a spurious perturbation solution as follows
\begin{equation}\label{gauge}
a=-iw\lambda,\quad p=e\lambda\psi_1,\quad \bar{p}=-e\lambda^*\psi_1,\quad q=\lambda\chi,\quad \bar{q}=-\lambda^*\chi.
\end{equation}
This spurious perturbation solution can be removed by setting $\lambda=0$, which can be further implemented by requiring $a=0$ on the horizon\footnote{The only one exception is the spurious mode with $w=0$. We care mainly about, however, the lowest lying mode in the lower complex frequency plane.}.

The boundary conditions \eqref{bdry} together with $a=0$ are naturally proposed with clear physical meaning. Whereas in the gauge invariant formalism, it is difficult to propose proper boundary conditions that exactly correspond to the gauge free boundary conditions that we need in the equations of dynamical evolution in the coexisting phase. This is one of the advantages of our method.

Finally, we cast the above linear perturbation equations and boundary conditions into the form $\mathcal{L}(w)v = 0$ with $v$ the perturbation fields evaluated at the grid points by the pseudo-spectral method. The quasi-normal frequencies are then obtained by the condition $\det[\mathcal{L}(w)] = 0$, which can be further identified by the density plot of $|{\det[\mathcal{L}(\omega)]'}/{\det[\mathcal{L}(\omega)]}|$ with the prime the derivative with respect to $\omega$. The relevant results are summarized as follows.

\begin{figure}[htbp]
\centering
\includegraphics[scale=0.43]{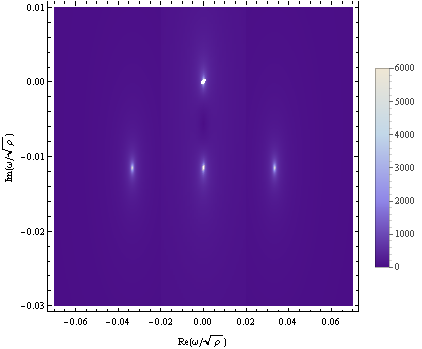}
\includegraphics[scale=0.46]{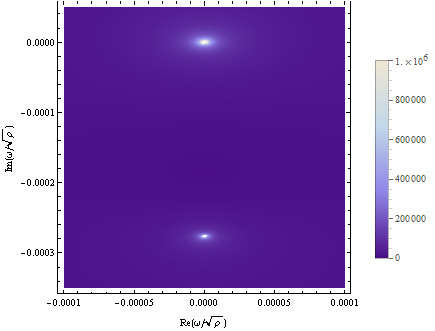}
\caption{ The left panel shows the QNMs with $\rho_d=3.99$. The QNMs siting on the origin and imaginary axis come from $\delta\psi_1$ coupled to $\delta A_t$, while the  QNMs lying symmetrically away from the imaginary axis come from $\delta\chi$. The right panel shows the QNMs with $\rho_m=8.44$.}\label{diacritical}
\end{figure}

When $\rho<\rho_1$, the system is in the normal phase and we thus have two sets of decoupled quasi-normal modes from $\delta\psi_1$ and $\delta\chi$, respectively. Both of the lowest lying modes are sitting symmetrically with respect to the imaginary axis. They migrate towards the origin with the increase of the charge density and the lowest lying modes from $\psi_1$ meet at the origin when $\rho_1$ is reached, which indicates the occurrence of phase transition to Phase-I.

When $\rho_1<\rho<\rho_{c1}$, the system is in Phase-I and we still have two sets of decoupled quasi-normal modes, where one is from $\delta\psi_1$ coupled to $\delta A_t$, and the other is from $\delta\chi$. The lowest lying modes from $\delta\chi$ keep migrating with the increase of the charge density and meet at the origin when $\rho_{c1}$ is reached, signaling the phase transition to Phase-III.  Among the quasi-normal modes from $\delta\psi_1$ and $\delta A_t$, one mode is pinned at the origin as Goldstone mode, and the other travels down the imaginary axis as Higgs mode with the increase of the charge density, eventually  exceeded by the aforementioned lowest lying mode from $\delta\chi$ at a certain charge density $\rho_d$ as plotted in Figure~\ref{diacritical}.

When $\rho_{c1}<\rho<\rho_{c2}$, the system is in the coexistent phase and we only have one set of quasi-normal models from the coupled $\delta\psi_1$, $\delta A_t$, and $\delta\chi$. The lowest lying mode sits right at the imaginary axis. With the increase of the charge density, this mode travels down the imaginary axis from the origin till a certain charge density $\rho_m$ and then travels up to the origin as plotted in Figure~\ref{diacritical}.

When $\rho>\rho_{c2}$, the system is in Phase-II and we again have two sets of decoupled quasi-normal modes, where one is from $\delta\psi_1$ and the other is from $\delta\chi$ coupled to $\delta A_t$. The lowest lying modes from $\delta\psi_1$ migrate symmetrically away from the origin with the increase of the charge density. While the Higgs mode from $\delta\chi$ and $\delta A_t$ keeps traveling down the imaginary axis, eventually exceeded by the climbing-up modes, which sit symmetrically with respect to the imaginary axis.

{In addition, we ascertain the points of phase transition by seeing that the lowest lying QNM crosses the real axis, and confirm the results that have been accurately obtained from the nonlinear static solutions. In some sense, the QNM provides a quicker and more reliable method to fix the points of phase transition, since it only involves solving the linear equations of perturbation, while the Newton-Raphson method used to solve the nonlinear static equations is inevitably sensitive to initial approximations.}

As alluded to in the very beginning, suppose that the fully non-linear dynamical evolution ends up with one of the above phases, then the late time behavior is believed to be captured by the lowest lying quasi-normal modes on top of the final equilibrium state. To be more specific, if the final equilibrium state has a non-vanishing condensate $|\langle O_f\rangle|\ne0$, then the late time behavior is supposed  to be approximated by
\begin{align}
|\langle O_t\rangle|=|\langle O_f\rangle|+\delta e^{-iw t}+\delta^{*}e^{iw^* t}.
\end{align}
On the other hand, if the final condensate  $|\langle O_f\rangle|=0$, then the late time behavior can be captured by
\begin{align}
|\langle O_t\rangle|=|\langle O_f\rangle+\delta e^{-iw t}|=|\delta|e^{Im(w)t}.
\end{align}

In the next section, we shall check whether these thermodynamically stable phases are also dynamically stable at the fully non-linear level by the real time evolution and see whether the late time behavior can be captured by our lowest lying quasi-normal modes(QNMs).

\section{Real Time Evolution and Dynamical Stability}
\subsection{Initial data and evolution scheme}

Although the parameter space of $\rho$ is divided into five parts by points $\rho_1$, $\rho_2$, $\rho_{c1}$, and $\rho_{c2}$,
%ref.~\cite{sonner} has discussed the evolution and the QNM of the system when given a quench to it with one scalar field $\psi_2$, and they work under the back reaction background. By giving a periodical driven in the $A_x$ direction, ref.~\cite{tian} investigate the system with one scalar field $\psi_2$. If one set $A_x$ to be a constant in terms of time, the system will evolve with this kind of initial perturbation. Above all, we ignore the area $\rho\le\rho_2$ for there is at most one superconducting phase {\blue and the dynamical characteristics have been investigated in certain aspects}. Now
in what follows we shall focus only onto the regime where there can exist at least two superconducting phases, namely $\rho_2\le\rho\le\rho_{c1}$, $\rho_{c1}\le\rho\le\rho_{c2}$ and $\rho\ge\rho_{c2}$.

Because Phase-I and Phase-II are always the possible phases in the above three regions, we will take two different kinds of initial data in each region. The first case is that we only let $\psi_1$ under the superconducting phase while give $\chi$ a perturbation of Gaussian wave packets, and we call this Case-I. The second case is that we only let $\chi$ under condensation while give $\psi_1$ a perturbation of Gaussian wave packets, and we call this Case-II.

In each case, the perturbation of Gaussian wave packets takes the form as $s*e^{-30(z-0.5)^2}$ with $s=1,\;0.1\;\mbox{or}\;0.01$, which satisfies the source free boundary condition on the AdS boundary within our numeric error. In order to decrease the error during the long-time evolution, we do not evolve the dynamical equation \eqref{At} of $A_t$, but obtain $A_t$ by solving the constraint equation \eqref{constr}, combining the boundary condition $A_t^{'}=-\rho$ on the AdS boundary and $A_t=0$ on the horizon.

With the above fixed charge density, we resort to the pseudo-spectral method to discretize the spatial direction, while in the time direction we take the forth order Runge-Kutta method to perform the evolution.The relevant numerical results will be presented in the following sections.

\subsection{$\rho_2\le\rho\le\rho_{c1}$}
\begin{figure}[htbp]
\centering
\includegraphics[scale=0.52]{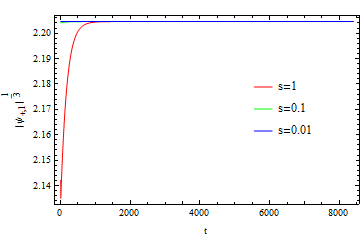}
\includegraphics[scale=0.52]{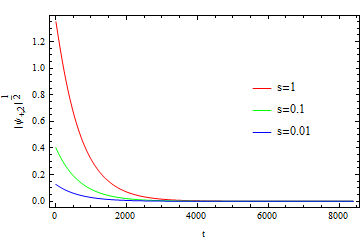}
\caption{The two figures show the Case-I with $\rho=7.5$ The left one shows the time evolution of condensation of the massless scalar field $\psi_1$ while the right one shows the time evolution of condensation of the scalar field $\chi$ during the same time period.}\label{751}
\end{figure}

\begin{figure}[htbp]
\centering
\includegraphics[scale=0.52]{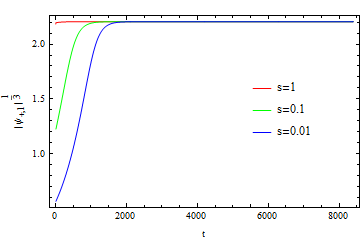}
\includegraphics[scale=0.52]{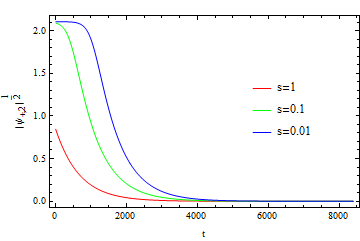}
\caption{The two figures show the Case-II with $\rho=7.5$. The left one shows the time evolution of condensation of the massless scalar field $\psi_1$  while the right one shows the time evolution of condensation of the scalar field $\chi$ during the same time period.}\label{752}
\end{figure}

Phase-I is thermodynamically stable in this region of $\rho$ and we take $\rho=7.5$ as an example. As shown in Figure~\ref{751},  the larger the perturbation strength is, the more time it takes for the system to approach Phase-I  in Case-I, while for Case-II, the system exhibits  an inverse behavior as the perturbation strengthens, which is demonstrated in Figure~\ref{752}.


\subsection{$\rho_{c1}\le\rho\le\rho_{c2}$}
\begin{figure}[htbp]
\centering
\includegraphics[scale=0.52]{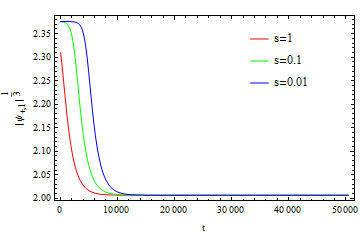}
\includegraphics[scale=0.52]{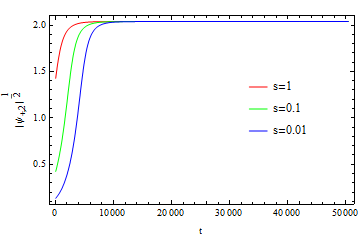}
\caption{The two figures show the Case-I with $\rho=8.5$. The left one shows the time evolution of condensation of the massless scalar field $\psi_1$ while the right one shows the time evolution of condensation of the scalar field $\chi$ during the same time period.}\label{851}
\end{figure}

\begin{figure}[htbp]
\centering
\includegraphics[scale=0.52]{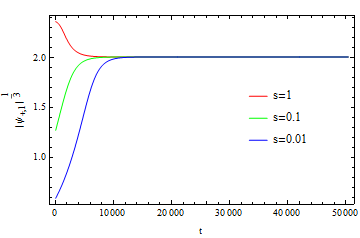}
\includegraphics[scale=0.52]{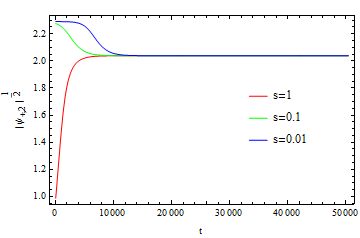}
\caption{The figures show the Case-II with $\rho=8.5$. The left one shows the time evolution of condensation of the massless scalar field $\psi_1$  while the right one shows the time evolution of condensation of the scalar field $\chi$ during the same time period.}\label{852}
\end{figure}

Phase-III is thermodynamically stable in this region of $\rho$ and we take $\rho=8.5$ as an example. As shown in Figure~\ref{851} for Case-I and Figure~\ref{852} for Case-II, the larger the perturbation strength is, the less time it takes for the system to approach Phase-III for both cases.


\subsection{$\rho\ge\rho_{c2}$}

\begin{figure}[htbp]
\centering
\includegraphics[scale=0.52]{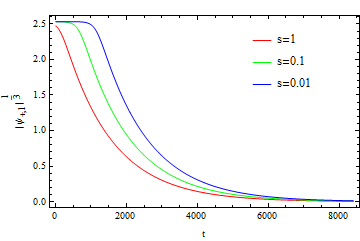}
\includegraphics[scale=0.52]{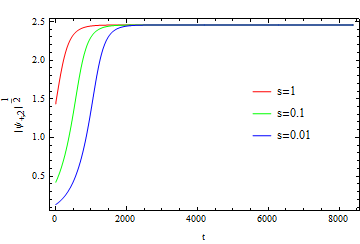}
\caption{The two figures show the Case-I with $\rho=9.5$. The left one shows the time evolution of condensation of the massless scalar field $\psi_1$  while the right one shows the time evolution of  condensation of the scalar field $\chi$ during the same time period.}\label{951}
\end{figure}

\begin{figure}[htbp]
\centering
\includegraphics[scale=0.52]{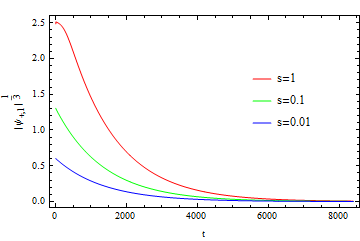}
\includegraphics[scale=0.52]{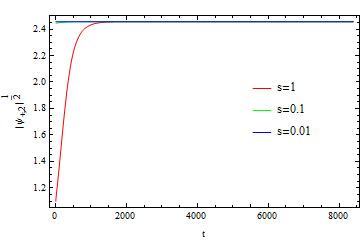}
\caption{The figures show the Case-II with $\rho=9.5$. The left one shows the time evolution of condensation of the massless scalar field $\psi_1$ while the right one shows the time evolution of condensation of the scalar field $\chi$ during the same time period where the evolution lines $s=0.1$ and $s=0.01$ are too close to distinguish.}\label{952}
\end{figure}

Phase-II is thermodynamically stable in this region of $\rho$ and we take $\rho=9.5$ as an example. As shown in Figure~\ref{951} for  Case-I, the larger the perturbation strength, the less time it takes for the system to approach Phase-II . For Case-II, the system exhibits  an inverse behavior, which is demonstrated in Figure~\ref{952}.


\subsection{Late time behavior towards the equilibrium state}

We see that the dynamical stability of the system is consistent with the thermodynamical stability in all the regions we are concerned with. But now we would like to turn to the detailed analysis of late time behavior of such a relaxation.

\begin{figure}[htbp]
\centering
\includegraphics[scale=0.42]{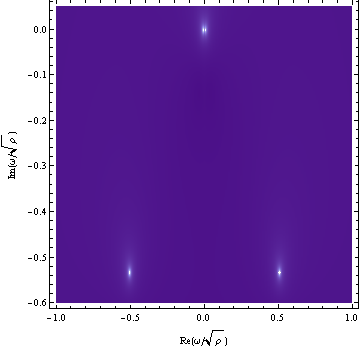}
\includegraphics[scale=0.45]{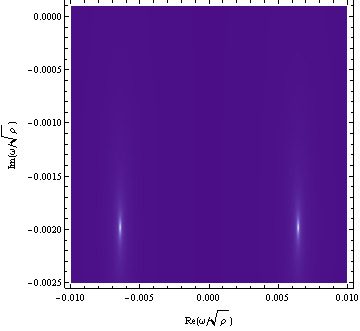}
\caption{The panels show the QNMs for the dying-out order $\chi$ with $\rho=7$. The left one shows the large scale situation with spurious $\omega=0$ mode at the origin. The extraordinary relaxation time implies that we should amplify the origin area to find the real lowest lying QNM, and the right panel shows the lowest lying QNMs we want.}\label{7}
\end{figure}

One of the remarkable characteristics of the evolution is that the relaxation time of the system is generically much longer than that of the simplest holographic system with only one scalar field.\footnote{The relaxation time of any system should diverge when tending to the critical points, so we focus on the cases that are far from criticality.} The relaxation time in such simplest system at $6.5<\rho<7.5$ is less than one hundred time units. However, it is several thousand time units here at $6.5<\rho<7.5$ (in the non-coexisting phase), so the difference is two orders of magnitude. This characteristic in return indicates us that the lowest lying QNM must be very close to the real axis (see Figure~\ref{7} for example), which would otherwise be confused with the mode $w=0$. It is interesting to investigate the possible physical mechanism behind such long relaxation time in this model.

\begin{figure}[htbp]
\centering
\includegraphics[scale=0.52]{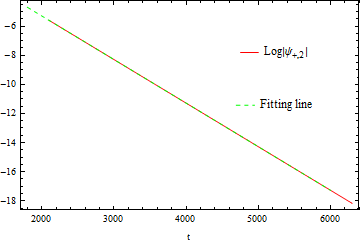}
\includegraphics[scale=0.52]{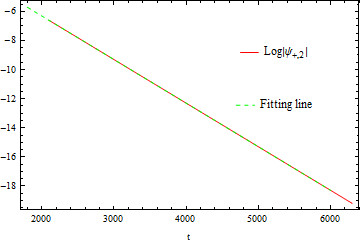}
\caption{The fitting plot for $\chi$ at $\rho=7.5$. In the left panel, the red line is the time evolution of logarithm of condensation of $\chi$ towards the final state for Case-I, and the green line is the fitting line with $Log(|\psi_{+,2}|)=0.692259-(0.0010934\sqrt{\rho})t$, so the decay frequency is $-0.0010934\sqrt{\rho}$. The right panel is the fitting for Case-II, and the fitting line is $Log(|\psi_{+,2}|)=-0.322169-(0.0010934\sqrt{\rho})t$, so the decay frequency is also $-0.0010934\sqrt{\rho}$. This decay frequency is consistent with the lowest lying QNMs in the right panel of Figure~\ref{s7}.}\label{753}
\end{figure}

\begin{figure}[htbp]
\centering
\includegraphics[scale=0.42]{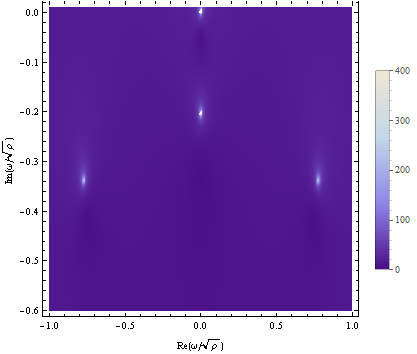}
\includegraphics[scale=0.45]{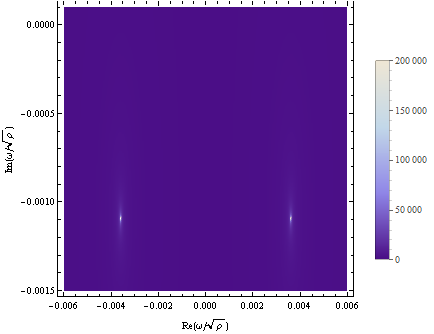}
\caption{The density plot of QNMs for $\rho=7.5$. The left panel shows the QNMs from $\delta\psi_1$ coupled to $\delta A_t$, and the lowest lying QNM sits on the imaginary axis with  $w=-i0.20\sqrt{\rho}$. The right panel is the QNMs from $\delta\chi$, and the imaginary part of the lowest lying mode is $-0.00109\sqrt{\rho}$.}\label{s7}
\end{figure}

Without loss of generality, we focus only onto the case of $s=1$. As fitted in Figure~\ref{753} for $\chi$ at $\rho=7.5$, there is an exponential decay with the decay frequency consistent with the lowest lying QNMs for $\chi$, which is  obtained by the aforementioned density plot in the right panel of Figure~\ref{s7}.

We find the late time behavior  of $\psi_1$ is also controlled by the lowest lying QNMs at  $\rho=9.5$, where the decay frequency is obtained as  $-0.000718787\sqrt{\rho}$. The late time behavior towards the coexisting phase is controlled by an exponential decay as well, with decay frequency $-0.00025727\sqrt{\rho}$ for $\rho=8.5$, which is consistent with its lowest lying QNM.

As we see, so far the late time behavior can be well captured by the lowest lying QNMs from the linear perturbation theory. But we find the non-linear effect starts to play a role in the late time behavior for $\psi_1$ in Phase-I and $\chi$ in Phase-II.

\begin{figure}[htbp]
\centering
\includegraphics[scale=0.52]{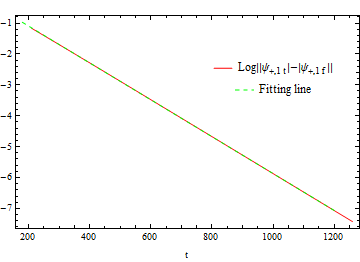}
\includegraphics[scale=0.52]{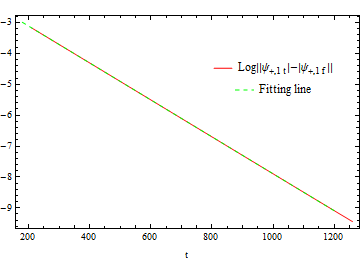}
\caption{The fitting plot for $\psi_1$ at $\rho=7.5$. The left panel for Case-I with the fitting line $Log(||\psi_{+,1t}|-|\psi_{+,1f}||)=0.101703-(0.0021803\sqrt{\rho})t$ ($\psi_{+,1t}$ denotes the condensation of $\psi_1$ during evolution, and $\psi_{+,1f}$ denotes the condensation in final equilibrium state). The right panel is for Case-II with the fitting line $Log(||\psi_{+,1t}|-|\psi_{+,1f}||)=-1.91539-(0.00218354\sqrt{\rho})t$. The decay frequency for $\psi_1$ turns out to be the double of  that for $\chi$. }\label{775}
\end{figure}

To be concrete, we denote the ratio of the absolute value of the imaginary part of the lowest lying QNM from $\delta\psi_1$ coupled to $\delta A_t$ over that from $\delta\chi$ as $r_1$ for Phase-I, and the ratio of the absolute value of the imaginary part of the lowest lying QNM from $\delta\chi$ coupled to $\delta A_t$ over that from $\delta\psi_1$ as $r_2$ for Phase-II. In the case of $\rho=7.5$, we can see from Figure~\ref{s7} that the lowest lying QNM from $\delta\psi_1$ coupled to $\delta A_t$ sits at the imaginary axis, and $r_1$ is around 200. But as fitted in Figure~\ref{775}, the real decay frequency for $\psi_1$ turns out to be the double of that for $\chi$ rather than its own lowest lying QNM frequency. In the case of $\rho=9.5$, the decay frequency for $\chi$ is $-0.00144\sqrt{\rho}$ which is also the double of that for $\psi_1$ mentioned before although the lowest lying QNM frequency for $\chi$ has the ratio $r_2=500$ or so.

Such a double relation is believed to come from the very non-linear character of our full equations of motion. To be more specific, due to the above large ratio $r_1$ or $r_2$, the lowest lying QNM for the non-vanishing order has died away long before the late time evolution in consideration. Then it follows from our non-linear equations of motion that such a double relation is generated by taking the lowest lying QNMs for the dying-out order as a source. This can also be regarded as the competition between the two orders, namely the dying-out order always tends to keep the other from relaxing towards the equilibrium state. Actually further numerical investigation indicates that this double relation occurs when $r_1$ or $r_2$ is bigger than $6$. Beyond this regime, such a double relation becomes obscure and the situation becomes more complicated due to the non-linear effect.

\section{Conclusion}
In this paper, we investigate the dynamical stability of the two band holographic superconductor (the BHMRS model). In passing, we find it more convenient (if not inevitable) for one to calculate the QNMs by gauge fixing the gauge dependent formalism. Actually, this method is especially adapted for models with more than one scalar fields and for direct comparison with the late time behavior of the dynamical evolution as we performed later in this paper. As a result, the thermodynamically stable phase is also dynamically stable at the linear level.

Furthermore, by evolving the system with different Gaussian wave packets as perturbations at the initial time, we find the dynamical stability of the system is also consistent with the thermodynamical stability at the fully non-linear level. We also analyze the late time behavior towards the final equilibrium state, and find it consistent with the lowest lying QNM in the coexisting phase as well as the lowest lying QNM for the field whose order dies out in the single ordered phases, while the late time behavior for the field whose order parameter does not vanish in the single ordered phases obeys a double relation or behaves in a much complicated way because of the non-linear effect induced competition between the two orders. In this sense, not only does the competition exhibit its existence in the phase diagram, but also does leave a footprint in the dynamical evolution. In addition, to our knowledge, this footprint is supposed to be the first counter example to the general belief that the late time behavior towards a final stable state can be captured by the lowest lying QNM obtained from the linear perturbation theory.

In comparison to the simplest holographic model with only one scalar field, the long relaxation time of the BHMRS model far from criticality is unexpected and interesting. This phenomenon may need further investigation, in particular for various other models with more than one order parameter.

\acknowledgments
We thank Weijia Li for his helpful discussion on the relevant numerics related to quasi-normal modes. We also thank Li Li and Zhangyu Nie for their valuable discussion on the free energy. In addition, we are grateful to Chao Niu and Xiaoning Wu for their useful discussions and comments on our manuscript. This work is supported by the Natural Science Foundation of China under Grant No. 11475179. Shan-Quan Lan is supported by the Natural Science Foundation of China under Grant No. 11235003. Hongbao Zhang is supported in part by the Belgian Federal Science Policy Office through the Interuniversity Attraction Pole P7/37, by FWO-Vlaanderen through the project G020714N, and by the Vrije Universiteit Brussel through the Strategic Research Program "High-Energy Physics". He is also an individual FWO Fellow supported by 12G3515N.

\section*{ Appendix: Derivation for the static equations}
Rewriting static equations \eqref{ste1}-\eqref{ste2} by replacing $\psi_1=|\psi_1|e^{ie\theta_1}=\phi_1e^{ie\theta_1}$, $\chi=|\chi|e^{i\theta_2}=\phi_2e^{i\theta_2}$, we obtain six equations. Besides  \eqref{static1}, \eqref{static2} and \eqref{static3}, the other three are:
\begin{align}\label{phi1}
A_t\phi_1^{\prime}-\frac{1}{z}A_t\phi_1+\frac{1}{2}A_t^{\prime}\phi_1+f\phi_1^{\prime}\theta_1^{\prime}+\frac{f}{2}\phi_1\theta_1^{\prime\prime}+(\frac{f^{\prime}}{2}-\frac{f}{z})\phi_1\theta_1^{\prime}=0,
\end{align}
\begin{align}\label{phi2}
A_t\phi_2^{\prime}+\frac{1}{2}A_t^{\prime}\phi_2+f\phi_2^{\prime}\theta_2^{\prime}+\frac{f}{2}\phi_2\theta_2^{\prime\prime}+\frac{f^{\prime}}{2}\phi_2\theta_2^{\prime}=0,
\end{align}
\begin{align}\label{phi3}
\frac{e^2\phi_1^2}{z^2}(f\theta_1^{\prime}+A_t)+\phi_2^2(f\theta_2^{\prime}+A_t)=0.
\end{align}

Multiplying the equation \eqref{phi1} by $\phi_1$, we obtain
\begin{align}
(\frac{A_t}{2}\phi_1^2+\frac{f\theta_1^{\prime}}{2}\phi_1^2)^{\prime}=\frac{2}{z}(\frac{A_t}{2}\phi_1^2+\frac{f\theta_1^{\prime}}{2}\phi_1^2),
\end{align}
which gives rise to
\begin{align}
\frac{A_t}{2}\phi_1^2+\frac{f\theta_1^{\prime}}{2}\phi_1^2=Cz^2.
\end{align}

Note that we have $f=0$ at the horizon $z=1$ as well as $A_t=0$ there by \eqref{phi3}, so we obtain $C=0$, which further implies
\begin{align}\label{static41}
A_t+f\theta_1^{\prime}=0.
\end{align}
Similarly we can have
\begin{align}\label{static51}
A_t+f\theta_2^{\prime}=0.
\end{align}
These two equations make \eqref{phi3} automatically satisfied.


\begin{thebibliography}{99}
\bibitem{gub}
S.S. Gubser and I. Mitra, \emph{Instability of charged black holes in anti-de Sitter space}, [\href{http://arxiv.org/abs/hep-th/0009126v1}{arXiv:0009126}].

\bibitem{gub1}
S.S. Gubser and I. Mitra, \emph{The evolution of unstable black holes in anti-de Sitter space}, JHEP
{\bf 0108}, 018 (2001) [\href{http://arxiv.org/abs/hep-th/0011127}{arXiv:hep-th/0011127}].

\bibitem{he1}
J.J. Friess, S.S. Gubser and I. Mitra, \emph{Counter-examples to the correlated stability conjecture}, Phys. Rev. D
{\bf 72}, 104019 (2005) [\href{http://arxiv.org/abs/1007.3480v3}{arXiv:1007.3480}].

\bibitem{he2}
A. Buchel, \emph{A holographic perspective on Gubser-Mitra conjecture}, Nucl. Phys. B
{\bf 731}, 109-124 (2005) [\href{http://arxiv.org/abs/hep-th/0507275}{arXiv:0507275}].

\bibitem{he3}
H.S. Reall, \emph{Classical and thermodynamic stability of black branes}, 	Phys. Rev. D
{\bf 64}, 044005 (2001) [\href{http://arxiv.org/abs/hep-th/0104071}{arXiv:0104071}].

\bibitem{he4}
T. Harmark, V. Niarchos, N.A. Obers, \emph{Instabilities of near-extremal smeared branes and the correlated stability conjecture}, JHEP
{\bf 0510}, 045 (2005) [\href{http://arxiv.org/abs/hep-th/0509011}{arXiv:0509011}].

\bibitem{he5}
T. Hirayama, G. Kang and Y. Lee, \emph{Classical stability of charged black branes and the Gubser-Mitra conjecture}, Phys. Rev. D
{\bf 67}, 024007 (2003) [\href{http://arxiv.org/abs/1007.3480v3}{arXiv:1007.3480}].

\bibitem{he6}
U. Miyamoto and H. Kudoh, \emph{New stable phase of non-uniform charged black strings}, JHEP
{\bf 0612}, 048 (2006) [\href{http://arxiv.org/abs/gr-qc/0609046}{arXiv:0609046}].

\bibitem{h3}
X.-H. Ge, Y. Ling, C. Niu and S.-J. Sin, \emph{Holographic transports and stability in anisotropic linear axion model}, [\href{http://arxiv.org/abs/1412.8346}{arXiv:1412.8346}].

\bibitem{he}
P. Basu, J. He, A. Mukherjee, M. Rozali and H.-H. Shieh, \emph{Competing holographic orders}, JHEP
{\bf 1010}, 092 (2010) [\href{http://arxiv.org/abs/1007.3480v3}{arXiv:1007.3480}].

\bibitem{le1}
A.J. Leggett, \emph{Inequalities, instabilities, and renormalization in metals and other fermi liquids},  \href{http://www.sciencedirect.com/science/article/pii/0003491668903047}{Annals of Physics {\bf 46}, 76-118 (1968)}.

\bibitem{le3}
Suk Bum Chung, Hendrik Bluhm and Eun-Ah Kim, \emph{Stability of Half-Quantum Vortices in $p_x$+i$p_y$ Superconductors}, \href{http://journals.aps.org/prl/abstract/10.1103/PhysRevLett.99.197002}{ Phys. Rev. Lett. {\bf 99}, 197002 (2007)}.

\bibitem{julien}
J. Garaud, K.A.H. Sellin, J. J$\ddot{a}$ykk$\ddot{a}$ and E. Babaev, \emph{Skyrmions induced by dissipationless drag in U(1)xU(1) superconductors}, Phys. Rev. {\bf B 89}, 104508 (2014) [\href{http://arxiv.org/abs/1307.3211}{arXiv:1307.3211}].

\bibitem{lili}
R.-G. Cai, L. Li, L.-F. Li, Y.-Q. Wang, \emph{Competition and Coexistence of Order Parameters in Holographic Multi-Band Superconductors}, JHEP {\bf 1309}, 074 (2013) [\href{http://arxiv.org/abs/1307.2768v2}{arXiv:1307.2768}].

\bibitem{leaver}
E.W. Leaver, \emph{Quasinormal modes of Reissner-Nordstr\"{o}m black holes}, Phys. Rev. {\bf D 41}, 2986 (1990).

\bibitem{amado}
I. Amado, M. Kaminski and K. Landsteiner, \emph{Hydrodynamics of holographic superconductors}, JHEP {\bf 0905}, 021 (2009) [\href{http://arxiv.org/abs/0903.2209v3}{arXiv:0903.2209}].
\end{thebibliography}
\end{document}